\documentclass{edm_article}

\usepackage{url}
\usepackage{lipsum}
\usepackage{booktabs}
\usepackage{multirow}
\usepackage{graphicx}
\usepackage{tabularx}
\usepackage{tabularx}
\usepackage{caption}
\usepackage{mdframed}
\usepackage[export]{adjustbox}
\usepackage{amsmath}
\usepackage{enumitem}
\usepackage{float}
\usepackage{placeins}
\usepackage{afterpage}
\usepackage{titlesec}
\usepackage{xcolor}
\usepackage{subcaption}

\begin{document}

\title{CHOP: Integrating ChatGPT into \\EFL Oral Presentation Practice}


\numberofauthors{4}
\author{
\alignauthor Jungyoub Cha\\
       \affaddr{KAIST}\\
       \affaddr{South Korea}\\
       \email{jungyoub.cha@kaist.ac.kr}
\alignauthor Jieun Han\\
       \affaddr{KAIST}\\
       \affaddr{South Korea}\\
       \email{jieun\_han@kaist.ac.kr}
\alignauthor Haneul Yoo\\
       \affaddr{KAIST}\\
       \affaddr{South Korea}\\
       \email{haneul.yoo@kaist.ac.kr}
\and
\alignauthor Alice Oh\\
       \affaddr{KAIST}\\
       \affaddr{South Korea}\\
       \email{alice.oh@kaist.edu}
}

\maketitle

\begin{figure*}[ht!]
    \centering
    \includegraphics[width=\textwidth]{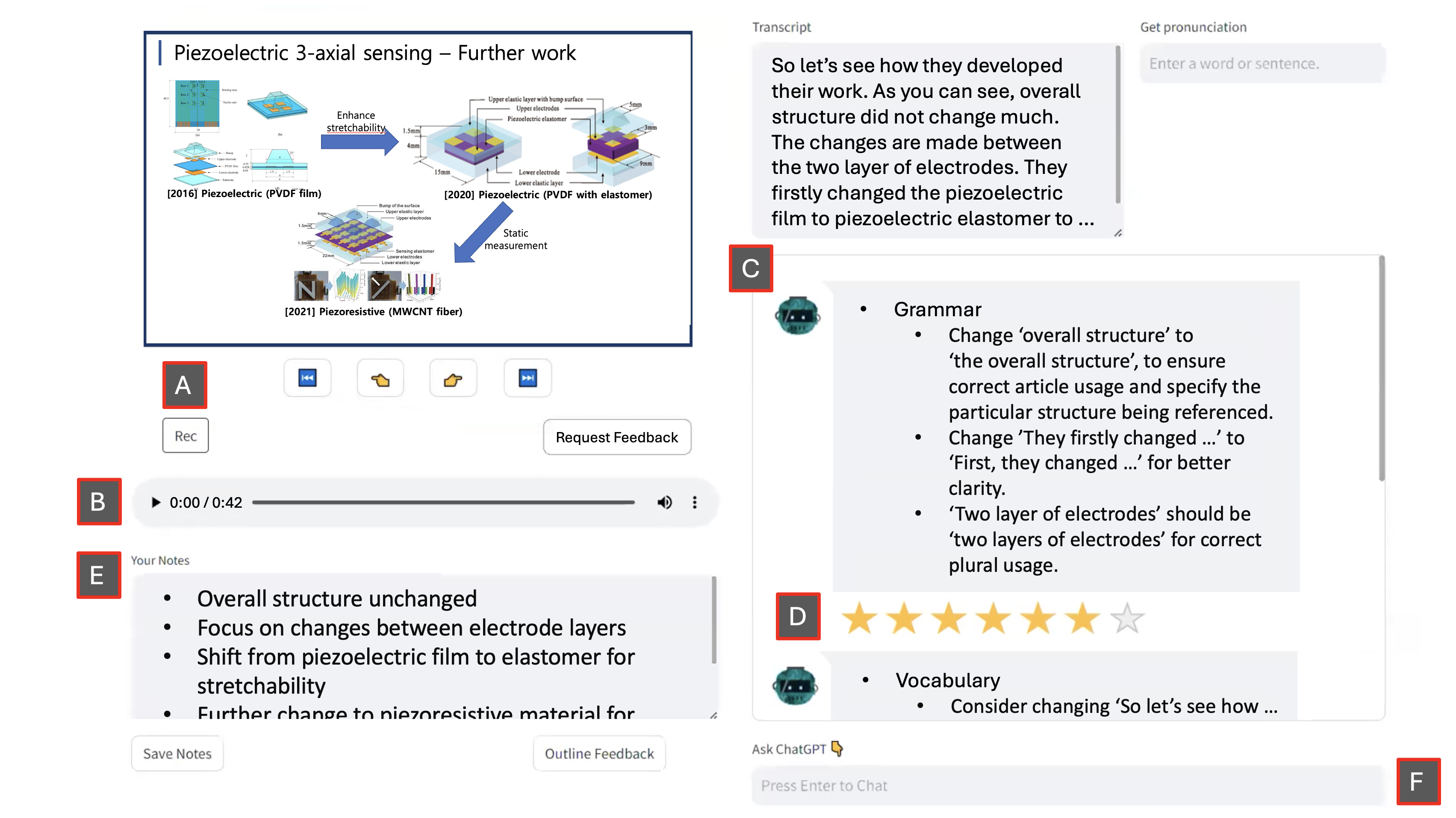}
    \caption{A screenshot of the default practice page on our platform. The student starts recording and can navigate through slides at (A) to practice any segment. They can playback the rehearsal audio at (B). Upon request, ChatGPT's feedback for each criterion is displayed at (C), along with the transcript of the rehearsal. The student is prompted to rate ChatGPT's every feedback in a 7-point Likert scale at (D). Then, the student can revise his notes at (E) and ask follow-up questions to ChatGPT at (F).}
    \label{fig:ui_screenshot}
\end{figure*}

\begin{abstract}

English as a Foreign Language (EFL) students often struggle to deliver oral presentations due to a lack of reliable resources and the limited effectiveness of instructors' feedback. Large Language Model (LLM) can offer new possibilities to assist students' oral presentations with real-time feedback. This paper investigates how ChatGPT can be effectively integrated into EFL oral presentation practice to provide personalized feedback. We introduce a novel learning platform, CHOP (ChatGPT-based interactive platform for oral presentation practice), and evaluate its effectiveness with 13 EFL students. By collecting student-ChatGPT interaction data and expert assessments of the feedback quality, we identify the platform’s strengths and weaknesses. We also analyze learners' perceptions and key design factors. Based on these insights, we suggest further development opportunities and design improvements for the education community. 

\end{abstract}



\section{Introduction}
\label{sec:intro}
Oral presentation skills are crucial for English as a Foreign Language (EFL) students to develop their overall communication skills \cite{ati-2022-use, brooks2015using} and prepare for academic and professional success \cite{algouzi2023strengthening}. Meanwhile, EFL students struggle to give effective oral presentations due to speech anxiety, trouble adapting information to spoken English, insufficient vocabulary repertoire and grammar knowledge, and mispronunciation \cite{hanifa2018insight, king-2002-preparing}. They face further difficulties due to a scarcity of effective teaching resources \cite{kho2023overcoming, seraj2021systematic} and the inadequacy of traditional teacher-centered approaches in meeting the dynamic needs of oral presentations \cite{aziz2022oral, tareen2023investigating}. This highlights the need to explore more interactive, student-centered methods that provide reliable and scalable assistance for oral presentation practice.

ChatGPT\thinspace\footnote{\label{chatgpt}\url{https://chat.openai.com/}}-assisted tools have the potential to enhance learning experiences for EFL students by providing personalized feedback \cite{hong2023impact, lo2023impact}. However, it is necessary to explore specific integration designs that take EFL learners' preferences and perceptions into account. While use cases of ChatGPT have been studied for writing education \cite{escalante2023ai, han2023recipe, meyer2024using} and conversational speaking practice \cite{hayashieffectiveness}, its role in oral presentation practice remains an open question.

In this study, we explore how to effectively integrate ChatGPT into oral presentation practice for EFL students. First, we conduct a focus group interview with five EFL students to understand their needs and preferences. Using these insights, we develop CHOP, a \textbf{CH}atGPT-integrated platform to assist with \textbf{O}ral \textbf{P}resentation practice by providing feedback on users' rehearsals. We test our platform with 13 students, collecting interaction data and students' post-survey responses about their experience, then have experts evaluate the quality of the generated feedback. By analyzing the collected data, we identify the potential learning effects as well as the strengths and weaknesses of the platform. We also highlight the factors that may affect the quality and learners' perceptions of the feedback, suggesting design improvements for further educational applications.

The main contributions are as follows:
\begin{itemize}
    \item We introduce CHOP, an interactive platform that uses ChatGPT to provide personalized feedback to EFL students for their oral presentation practice. 
    \item We investigate factors that affect the quality of feedback and learners' perceptions.
    \item We propose UI/design improvements to address these issues, investigating further development of our platform.
\end{itemize}



\section{Related Work}
\subsection{Oral Presentations in EFL Education}

Previous studies have explored enhancing EFL students' presentation skills through various methods. Workshops and interventions have been used to increase practice opportunities and self-confidence \cite{algouzi2023strengthening, rohman2023students}. Previous work has also explored the integration of technology into educational settings. Specifically, \cite{gokturk2016examining} assessed the impact of digital video recordings on EFL learners' oral performance, while \cite{shih2010blended} evaluated the effects of combining video-based blogs with conventional classroom instruction on public speaking skills. \cite{algouzi2023strengthening} also examined the use of Blackboard, an online learning management system, to improve students’ oral presentation skills. We extend this line of work and explore a solution that can provide personalized feedback to learners in real-time.

\subsection{Generative AI in Education}
ChatGPT, a large language model-based chatbot powered by OpenAI, has significantly advanced language learning \cite{lo2023impact, rahman2023chatgpt}. Its ability to understand complex text nuances and generate real-time feedback has been used in writing education to provide evaluation scores and feedback on essays \cite{barrot2023using, han2023fabric, whalen2023chatgpt}. 
Large language models have also been used for speaking education. For instance, ChatGPT has been used as a conversational partner in training tools to improve real-world conversation skills of learners \cite{hayashieffectiveness}. Educational applications, such as Duolingo\thinspace\footnote{\label{Duolingo}\url{https://blog.duolingo.com/duolingo-max/}} and Khan Academy\thinspace\footnote{\label{khanacademy}\url{https://www.khanacademy.org/}}, have incorporated large language models into their platform to provide further explanation and rationale to learners. Our work focuses on using ChatGPT to provide holistic feedback on users' oral performance according to a specific rubric. Furthermore, we shed light on oral presentations rather than casual conversations, which require deep contemplation of unique aspects in oral presentations, including but not limited to the content and organization of the presentation. To the best of our knowledge, our work firstly explores how ChatGPT can be integrated into oral presentation practice to provide holistic feedback on presentation rehearsals.

\section{Platform Design}
\subsection{Preliminary Analysis}
We conducted a focus group interview with five EFL students in South Korea to gain deeper insights into their need for oral presentation assistance. The interview details are in Appendix \ref{sec:appendix_fgi}. The main challenges for EFL students when giving oral presentations are limited vocabulary, nervousness or lack of confidence, appropriate formality, and correct pronunciation. We also discover that they prefer direct, specific, and negative feedback, which aligns with the findings of \cite{han2023fabric, sidman2015academic}, alongside a balance between immediate and delayed responses. 

Given that PPT-based oral presentation can enhance EFL students' essential soft skills \cite{dobson2006assessment, hanifa2018insight}, we design our platform to support PPT-based oral presentation. Students can present on a free genre or topic using PPT for 5 to 15 minutes through our platform.

\subsection{Implementation}
Based on established presentation rubrics \cite{al2014speaking, al2010taking, king-2002-preparing} and the specific requirements of EFL students from the interview, we implement our platform to provide feedback across the following presentation criteria: Grammar, Vocabulary, Content, Organization, and Delivery. We structure the feedback style as corrective, direct, specific, and negative, following students' preferences from our preliminary analysis. 

We use the \texttt{gpt-4-turbo-preview} model with a temperature of 0 to generate rubric-based feedback. Details on the system prompt for ChatGPT are in Appendix \ref{sec:appendix_prompt_details}. In the system prompt (Table \ref{tab:system_prompt}), we provide ChatGPT’s role as a presentation aid, the presentation context (e.g., where and to whom the presentation is being presented, and summarized slide notes. We then utilize Whisper\thinspace\footnote{\label{whisper}\url{https://openai.com/index/whisper}} to transcribe rehearsal audio and integrate this into the feedback prompt (Table \ref{tab:feedback_prompt}), along with the presentation rubric, feedback style, and an example to ensure consistent format. For feedback on the delivery component, we use SpeechSuper\thinspace\footnote{\label{speechsuper}\url{https://www.speechsuper.com/}} to assess the rehearsal audio on elements, such as pace, fluency, and pronunciation. It outputs scores (0-100) on each element and word- and sentence-level pronunciation feedback. We provide all interim results to ChatGPT for feedback generation on delivery.

Our platform offers two feedback settings to the students: partial rehearsal feedback and full rehearsal feedback. 
Partial rehearsal feedback, a default setting of our platform (Figure \ref{fig:ui_screenshot}), allows students to practice any specific segments of their presentations and immediately receive feedback. According to previous studies, students prefer receiving feedback with less delay as it allows them to interpret the feedback while the context is still fresh in their working memory \cite{gamlo2019efl, quinn2014delayed}. For this setting, ChatGPT is prompted to address every visible error in the rehearsal transcript.

However, applying the same detailed feedback approach to longer rehearsals, i.e., full rehearsals, proved impractical due to the overwhelming specificity and volume of comments. To resolve this issue, we separately create a full rehearsal mode, where we prompt ChatGPT to aggregate common errors and emphasize the types of errors rather than every single instance. This metalinguistic feedback allows users to receive concise yet informative guidance that supports long rehearsals \cite{liang2023chatback, rassaei2012effects}. Furthermore, students with this full rehearsal feedback setting can receive feedback with greater delay after their entire rehearsal is completed. Such feedback has strengths as it prevents disruption of practice flow and utilizes a broader context for more comprehensive feedback \cite{olmezer2016types, quinn2014delayed}.


\subsection{Experimental Design}
We carry out an experiment spanning two weeks in which 13 EFL students utilize our platform for their oral presentation practice. The details on the students' backgrounds are in Appendix \ref{sec:appendix_students_info}. Throughout this period, we collect all interaction data, including rehearsal audio, ChatGPT-generated feedback, user ratings, chatlogs, and platform logs. We then conduct a post-survey to further our understanding of the users' experiences and satisfaction with the platform. Details on the post-survey are in Appendix \ref{sec:appendix_postsurvey_details}.


To assess the quality of the generated feedback, we recruit 10 English education professionals, each holding a Secondary School Teacher's Certificate (Grade II) for English Language, licensed by the Ministry of Education in Korea. They evaluate the feedback based on accuracy, level of detail, relevance, and helpfulness, using a 7-point Likert scale. Throughout the experiment, a total of 289 feedback samples were generated for each presentation criterion. We randomly select two partial rehearsals and one full rehearsal from each participant, resulting in 39 feedback samples to be evaluated. 
Each sample is assessed by three experts, taking the average of three as the final score.

\section{Experimental Results}
\begin{figure}[t]
    \centering
    \includegraphics[width=4.5cm]{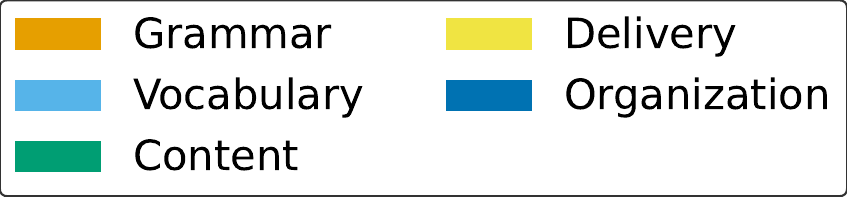}
    \includegraphics[width=\columnwidth]{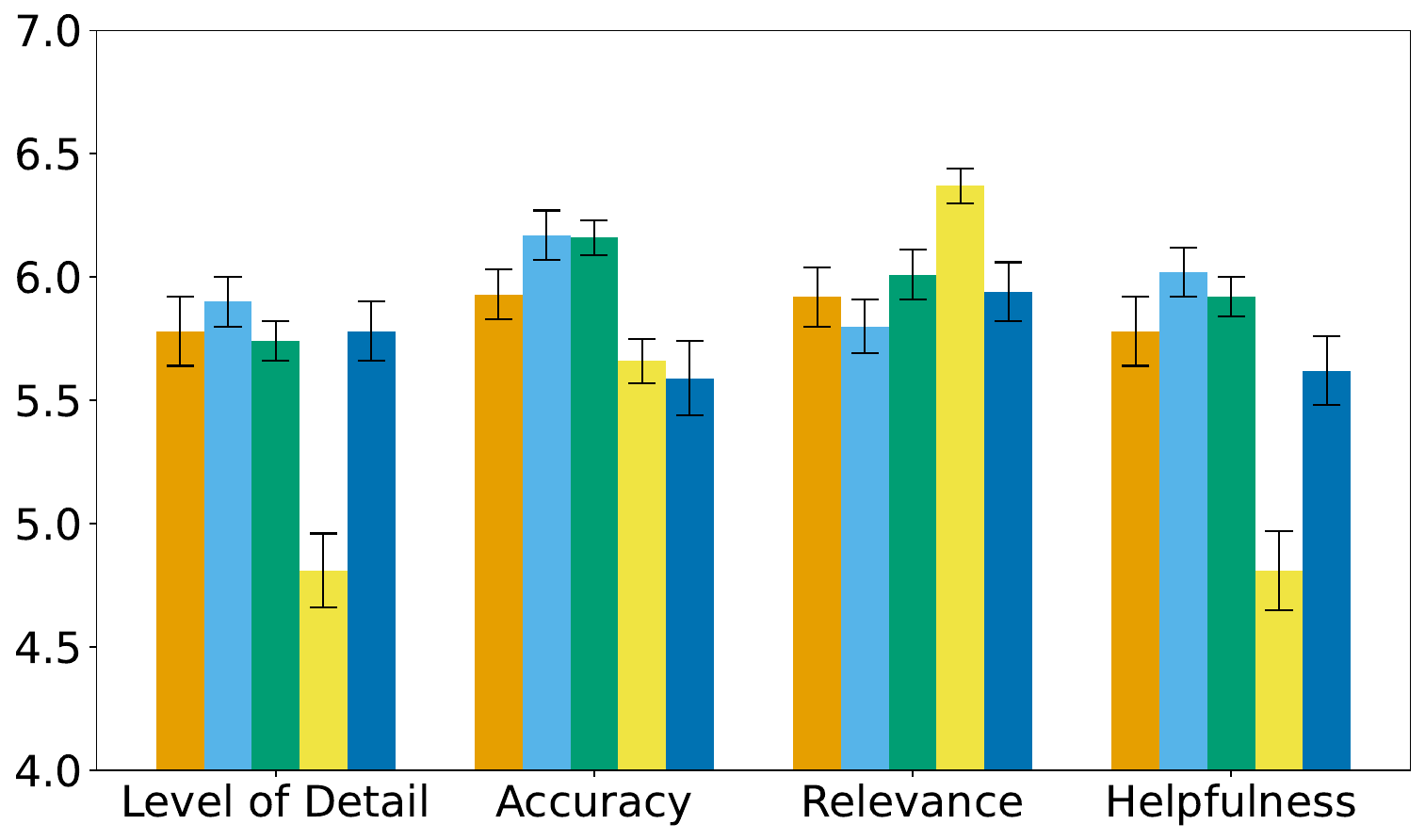}
    \caption{Evaluation results of feedback by presentation criteria in a 7-point Likert scale.}
    \label{fig:expert_eval_scores}
\end{figure}

In this section, we assess the effectiveness of our platform with feedback quality (\S\ref{sec:feedback_quality}) and learners' perceptions (\S\ref{sec:user_perception}). We then gain deeper insights by analyzing the influencing factors (\S\ref{sec:factors}).

\subsection{Feedback Quality Evaluation}
\label{sec:feedback_quality}

Figure \ref{fig:expert_eval_scores} shows English experts' evaluation results of the feedback quality. Across all presentation criteria, the feedback excels in accuracy and relevance, scoring a minimum of 5.66/7 and 5.8/7, respectively. This indicates that the platform is capable of accurately addressing errors that are relevant to the given presentation rubric. 

In particular, our platform is particularly effective in providing feedback on vocabulary, as perceived by both experts and students. It achieves the highest scores from experts for accuracy (6.17/7) and helpfulness (6.02/7). Students rate the feedback on vocabulary as the most helpful out of all criteria according to the post-survey (5.25/7). Students also rate 5.5/7 on average regarding the platform's effectiveness in enhancing their overall vocabulary skills.

Nonetheless, our feedback system shows its limitations in the delivery component, which received noticeably lower scores for level of detail (4.81/7) and helpfulness (4.81/7). Both experts and students report instances of ambiguity within the delivery feedback. Students (S2, S12) find it difficult to understand the exact rationale behind differing delivery scores across rehearsals. This is due to the limitations on the interpretability of black-box SuperSpeech API, which lacks details in the rationale of the scores. This indicates that the quality of feedback on delivery could be enhanced via improvements in the speech assessment tool’s transparency and capability to extract more granular information, such as pinpointing exact moments in the rehearsal when the pace was too fast.

Feedback on the content and organization components shows comparable scores to other criteria in terms of quality. Nevertheless, students (S6, S11) report that feedback on content and organization components requires greater effort to interpret and is more challenging to implement. This shows that the degree of difficulty and time needed to apply feedback can vary across presentation criteria, as perceived by learners. Thus, the feedback quantity or delivery mechanism should be flexibly adjusted for each criterion to prevent information overload.

\subsection{Learners' Perceptions}
\label{sec:user_perception}

We gain further insights into learners' perceptions of the platform and potential learning effects from the post-survey results. Students rate the platform's impact on improving their confidence and reducing nervousness at an average of 5.26 out of 7. S2 mentions that revising notes based on the feedback received increases their credibility, thereby boosting their confidence. This highlights the platform's potential to enhance learners' confidence, addressing a significant challenge for EFL students. Regarding the feedback's role in helping them recognize their strengths and weaknesses, students provide an average rating of 5.27 out of 7. S3 points out that a major benefit of the platform is making students aware of unrecognized habits. This suggests the platform's effectiveness in developing self-evaluation skills through a cycle of practice, feedback, and reflection. Furthermore, students rate how closely the full rehearsal mode resembles a real online presentation environment and its effectiveness in practicing online presentations, with average ratings of 4.92 and 6.33 out of 7, respectively. This indicates that the platform can potentially serve as an effective tool for online presentation practice, which is becoming increasingly common in today's environment.

\subsection{Influencing Factors for Feedback Quality and Learners' Perceptions}
\label{sec:factors}
While expert evaluations demonstrate the platform's potential to provide quality feedback, we further explore factors that affect feedback quality and learners' perceptions by analyzing usage patterns and post-survey results.
\subsubsection{Usage Patterns by Presentation Note Type}

\paragraph{Preliminary Analysis}
Presentation notes are crucial for helping students effectively deliver their content when giving oral presentations. Proficient English speakers, including EFL students, often utilize brief keyword notes to avoid reliance on a full manuscript \cite{hanifa2018insight}. This can facilitate more natural and fluent delivery while mitigating the common challenge of EFL students losing their train of thought due to difficulties in organizing ideas cohesively \cite{gani2015students, king-2002-preparing}. For our analysis, we categorize the presentation notes into two types: manuscripts, which contain the entire written script, and key points, comprising all non-manuscript notes of various levels of detail. In our experiment where participants could freely choose their preferred note type, the majority (nine students) chose manuscripts, while four students opted for key points. This is likely due to lower-level EFL students' struggles with forming fluent sentences from keywords alone \cite{al2014speaking} and the online presentation setting where direct audience interaction, like eye contact, is not required.

\paragraph{Feedback Response Analysis and Design Suggestions}
We examine how participants respond to feedback from ChatGPT by categorizing their immediate actions into two types: note revision and iterative rehearsal. Note revision refers to editing their notes based on feedback, while iterative rehearsal refers to rehearsing the slide again. 
\begin{table}[htb!]
\centering

\begin{tabular}{@{}ccc@{}}
\toprule
           & Note revision & Iterative rehearsal \\ \midrule
Key points & 25\%          & 75\%                \\
Manuscript & 100\%         & 89\%                \\ \bottomrule
\end{tabular}%
\caption{Percentage of students that display each type of feedback response at least once during their practice sessions. 
}

\label{tab:feedback_response_2}
\end{table}
\begin{table}[htb!]
\centering
\begin{tabular}{@{}lcc@{}}
\toprule
           & Note revision & Iterative rehearsal \\ \midrule
Key points & 0.24          & 0.31                \\
Manuscript & 0.76          & 0.17                \\ \bottomrule
\end{tabular}
\caption{Average number of feedback response per feedback}
\label{tab:feedback_response_1}
\end{table}

Table \ref{tab:feedback_response_2} shows key points users make note revisions (25\%) less compared to iterative rehearsals (75\%), while manuscript users show a balanced use of feedback responses. 
Table \ref{tab:feedback_response_1} indicates the average number of responses per feedback received. For each given feedback, key points users average more iterative rehearsals than note revisions, while manuscript users show the opposite trend. This discrepancy likely arises because it is difficult for key points users to integrate specific lexical or grammatical corrections into their notes to preserve their simplicity. The absence of a full manuscript makes precise revisions more difficult. For instance, in Figure \ref{fig:ui_screenshot}, given the grammar feedback \textit{``Change `They firstly changed ...' to `First, they changed' for better clarity.''}, the student cannot directly incorporate this into their notes as the addressed phrase is not included in the notes. Thus key points users instead resort to rehearsing the slide again. This finding is also supported by further analysis revealing that all instances of note revisions made by key points users were related to feedback on content only, unlike manuscript users who also made revisions related to vocabulary and grammar feedback. Manuscript users can fully prepare the content of their presentations by editing their notes without having to rehearse them repeatedly.


These observations underscore several implications for platform design and feedback mechanisms. The platform should offer flexible feedback tailored to each student's note type. For instance, in our current platform, key points users must remember specific vocabulary and grammar feedback and recall it from memory for future rehearsals. This is because of the difficulty of incorporating such detailed feedback into their notes. Therefore, we can enhance their learning experiences by providing feedback memory-efficiently, making it easier to incorporate into their subsequent rehearsals. This could involve using metalinguistic feedback to help learners engage more cognitively with the feedback, thereby enhancing its retention \cite{abdollahzadeh2016effect, amoli2020effect}. We also suggest UI enhancements such as a sticky notes feature for easier reference during subsequent rehearsals.

Manuscript users (S5, S11), on the other hand, show interest in separate features for note-specific feedback (e.g., ``\textit{Give me feedback on my notes.}''). This is because in our current platform, in order to receive additional feedback on their notes after revision, the student must rehearse it again, which is inefficient. This suggests that creating separate feedback features for the notes and rehearsal delivery could lead to more efficient learning for manuscript users. 

\begin{figure}[htbp]
    \centering
    \subfloat{\includegraphics[width=6cm]{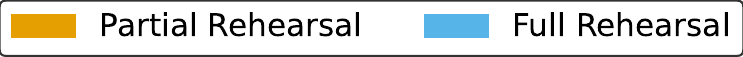}}
    \addtocounter{subfigure}{-1}
    \subfloat[\centering Platform Usage Experiment]{\includegraphics[height=4cm]{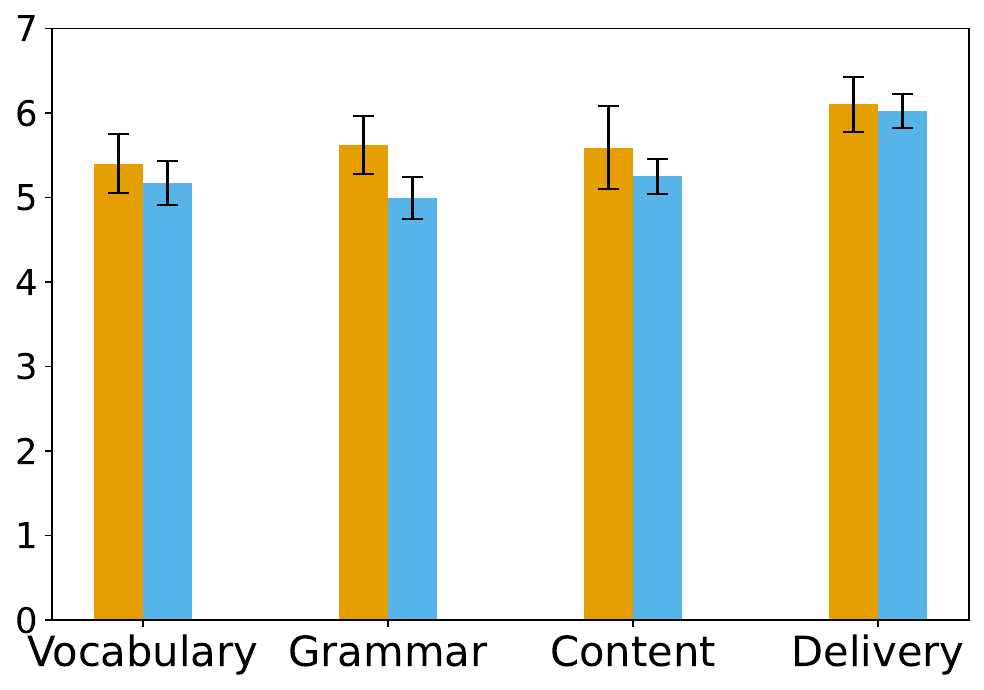}}\label{fig:partial_vs_full_1}
    \subfloat[\centering Post-survey]{\includegraphics[height=4cm]{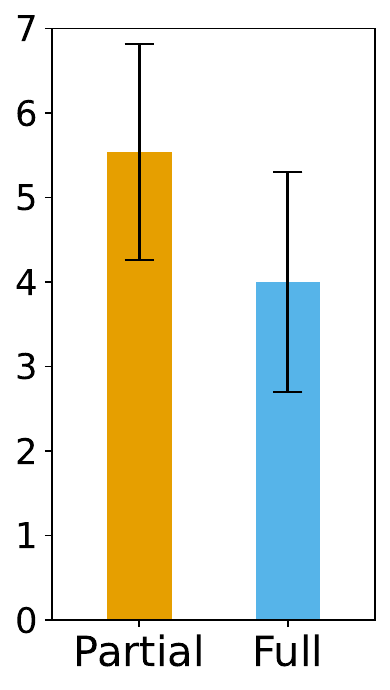}}\label{fig:partial_vs_full_2}
    \caption{Average user ratings of partial and full rehearsal feedback in 7-point Likert scales. (a) shows the average of ratings collected from the platform usage experiment. (b) shows the average ratings of overall satisfaction from the post-survey.}
    \label{fig:ratings}
\end{figure}

\subsubsection{Implications of Partial and Full Rehearsal Feedback Settings}
In this section, we share our findings on students' perceptions of the partial and full rehearsal feedback settings. First, we analyze how they utilize each mode throughout their practice sessions. We then draw on these perceptions to identify key design elements for both settings.

\paragraph{Usage patterns}
Usage patterns reveal that the predominant strategy among students (12 out of 13) is to first utilize partial rehearsals for detailed segment-by-segment refinement, then finish with full rehearsals for a comprehensive review. According to the post-survey, numerous students (S2, S4, S6, S9) recognize the value of partial rehearsal feedback in honing minor details thanks to its specificity, while full rehearsal feedback is praised for providing comprehensive insights, given its wide-ranging context (S2, S6). These findings underline the distinct yet complementary roles of each feedback mode.

\paragraph{Implications and Design Suggestions}
Challenges emerge concerning the context interpretation of full rehearsal feedback. Despite the comparable quality of feedback in both modes as evaluated by experts, students prefer partial over full rehearsal feedback in ratings from both the platform usage experiment and post-survey (Figure \ref{fig:ratings}). Further, students (S3, S6, S12) report difficulties in pinpointing the exact context of errors addressed in the full rehearsal feedback. Students struggle to apply the given feedback due to these challenges in understanding their context. This issue likely arises due to two factors: (1) the lengthy nature of the rehearsal audio which leads to difficulties in locating the error, and (2) the metalinguistic approach of addressing error types rather than each specific instance, initially incorporated to minimize feedback quantity.
Therefore, to improve the effectiveness of feedback for long rehearsals, the platform design should consider the appropriate amount and detail of feedback and ways to help the learner interpret the context of feedback. Potential UI enhancements include linking feedback to specific timestamps in rehearsal audio or transcripts, linking error types with their instances, and using various forms of media feedback, such as audio feedback.

Meanwhile, students (S3, S4, S11) report that partial rehearsal feedback sometimes contains too much trivial information. S3 claims that it even leads to distraction and impacts their confidence. This is likely due to the iterative nature of presentation practice on our platform, which involves multiple rehearsal and feedback cycles, implying that the feedback volume could affect the learner's practice flow. This suggests the need to carefully manage both the volume of feedback and the significance of errors, especially for partial rehearsals, which require more frequent iterations. For instance, dynamically adjusting the threshold of error severity for which feedback is provided, or providing feedback by order of significance could help ensure the feedback is more helpful and relevant to the learner.

\subsubsection{Repeated Feedback Requests}
Another issue emerges when a student requests feedback from the same slide multiple times. Students commonly practice this to rehearse the slide repeatedly and to request confirmatory feedback on the slide notes after making revisions. Students report instances where the feedback appears inconsistent with previous feedback (S2, S7) or remains unchanged despite revisions being applied (S4). Such inconsistencies could be mitigated by enhancing ChatGPT's context awareness of previous feedback for specific slides and optimizing prompting methods. This will enable ChatGPT to reference earlier feedback better and thereby offer more consistent feedback.

\subsubsection{Presentation Topic Complexity}
Students report a challenge in ChatGPT's inability to grasp the technical depth of presentation topics (S1, S6, S9, S11), all of their topics pertaining to science and engineering fields. Upon feedback sample analysis, the issue manifests in two ways: technical terms being incorrectly transcribed by Whisper and ChatGPT's inherent lack of expertise. For instance, ChatGPT may advise the student to replace an irreplaceable technical term (e.g., \textit{``The word `sampling' is overused. Consider replacing the second instance to `collecting' or `gathering.' ''}). Enhancements could include supplying the speech recognition model and LLM with additional expert knowledge and ensuring more accurate feedback in technical domains.


\section{Limitation}
In this section, we address the limitations of our work and opportunities for future improvement.
We test the platform on a relatively small sample of 13 EFL students, potentially affecting the generalizability of our findings.
Additionally, the platform only utilizes the user's presentation notes and rehearsal transcript to generate feedback using ChatGPT. In other words, it ignores other useful input signals, including but not limited to the raw rehearsal audio, PPT file, or the presenter's visual cues such as eye contact and gestures, which could offer valuable non-linguistic features. Incorporating such information to provide more comprehensive feedback could be an avenue for future improvement.
Lastly, as ChatGPT is a black-box language model, the feedback generated by our platform lacks transparency and clear rationale. We recognize the need for further research to develop models capable of producing more explainable feedback.

\section{Conclusion}

In this paper, we introduce CHOP, a ChatGPT-based interactive platform for EFL students, to provide personalized feedback on oral presentations.
CHOP provides students with improvements in confidence, vocabulary, self-assessment, and online presentation skills.
The feedback generated from our platform is perceived as helpful in vocabulary, content, grammar, and organization by both English experts and students.
CHOP addresses various students' needs in feedback mechanisms based on their practice patterns.
We identify design considerations for personalized feedback needs of students to effectively integrate ChatGPT into EFL education.
This study contributes to EFL education by providing insights into learners' perceptions and key design factors, which is invaluable for the further development of ChatGPT-based platforms for oral presentation practice.


\section{Acknowledgments}
This work was supported by Elice.

\bibliographystyle{abbrv}
\bibliography{sigproc}  
%
\clearpage
\appendix
\section{Details on Students}
\label{sec:appendix_students_info}
This section shows the demographic and educational backgrounds of the students who participated in the experiments. All students are from the Republic of Korea, and their average age is 24.5.
Table~\ref{tab:demographics} describes the demographic details of the participants.
\begin{table}[h]
\centering
\resizebox{\columnwidth}{!}{%
\begin{tabular}{@{}cccccc@{}}
\toprule
Student & Gender & Age & Degree & CEFR Level & Presentation note  \\ \midrule
S1      & Male   & 25  & Masters               & C1         & Manuscript             \\
S2      & Male   & 24  & Bachelors             & B2         & Manuscript             \\
S3      & Male   & 23  & Bachelors             & C1         & Key points             \\
S4      & Male   & 20  & Bachelors             & C1         & Key points             \\
S5      & Male   & 25  & Masters               & C1         & Manuscript             \\
S6      & Male   & 25  & Masters               & B2         & Manuscript             \\
S7      & Male   & 22  & Bachelors             & B2         & Manuscript             \\
S8      & Male   & 25  & Bachelors             & C1         & Manuscript             \\
S9      & Male   & 25  & Doctoral              & B2         & Manuscript             \\
S10     & Male   & 25  & Masters               & B2         & Manuscript             \\
S11     & Male   & 25  & Doctoral              & C1         & Manuscript             \\
S12     & Female & 29  & Masters               & C1         & Key points             \\
S13     & Male   & 25  & Bachelors             & C1         & Key points             \\ \bottomrule
\end{tabular}%
}
\caption{Demographic and educational backgrounds of students}
\label{tab:demographics}
\end{table}

\section{Details on Focus Group Interview}
\label{sec:appendix_fgi}
\subsection{Interview Design}
In this section, we describe the design of the preliminary focus group interview conducted to find the needs and preferences of EFL students with oral presentations. Students S1, S2, S3, S4, and S5 participated in the interview. The following paragraphs describe the questions used in the interview.

\textbf{Experience with Oral Presentations}
\begin{enumerate}
    \item How did you prepare for your oral presentations?
    \item What do you think are the key elements that contribute to an effective oral presentation?
    \item What challenges did you encounter while preparing for and delivering oral presentations?
    \item How did you attempt to overcome these challenges?
\end{enumerate}

\textbf{Feedback Preferences}
\begin{enumerate}
    \item Below are feedback samples with different styles: one is direct and the other is indirect. Which do you prefer and why?
    \item Below are feedback samples with different styles: one is positive and the other is negative. Which do you prefer and why?
    \item Below are feedback samples with different styles: one is vague and the other is specific. Which do you prefer and why?
    \item Below are feedback samples with different styles: one is straightforward and the other is polite. Which do you prefer and why?
    \item Below are feedback samples with different styles: one is immediate and the other is delayed. Which do you prefer and why?
\end{enumerate}



\subsection{Findings}
\subsubsection{Difficulties Encountered in Oral Presentations}
Students face various difficulties in preparing for oral presentations. S1, S3, and S4 state that a common issue is the overuse of filler words (e.g., ``however'' and ``that'') when presenting. This results from limited vocabulary, nervousness leading to unexpected delivery errors (e.g., speaking too fast and mispronouncing words), and difficulties in composing lengthy, well-structured sentences. S2, S5 also struggle with choosing appropriate language that matches the formality of the presentation context. When preparing their script, due to a lack of expert resources or reference materials, S2, S3, and S4 have trouble determining the appropriateness of their word choice, sentence structure, and overall script quality. S3 and S4 claim that feedback from translation tools often appears unnatural, yet they have no reliable means to verify or correct this.

\subsubsection{Approaches to Address Challenges}
S1, S2, and S5 address these issues by reviewing recordings of their own presentations and learning from them. They watch online videos to understand native and natural expressions. S4 and S5 create synonym banks from blog posts to enhance their vocabulary. However, they note that these resources are not always reliable as they are often produced by other non-native English speakers. To compensate for the lack of expert advice, S2, S3, and S4 consult peers who are more proficient in English. While S2 and S4 have tried using ChatGPT, they consider it time-consuming and ineffective, as it requires detailed explanations of the presentation context to get useful feedback.

\subsubsection{Criteria for an Effective Presentation}
We asked students what they perceive as an effective presentation and what it consists of. S1 and S2 highlight delivery aspects to be important factors, such as natural pronunciation and maintaining a fluent pace. S2 and S3 believe a good presentation goes beyond correct pronunciation or delivery; it must convey information effectively and engagingly to the audience. Unlike essays, presentation scripts do not have to be fancy; instead, they should be clear and concise.

\subsubsection{Desired Assistance}
S4 and S5 expressed their need for comprehensive assistance in writing the script and detailed feedback on delivery aspects like pace, accent, and body language. S1 and S3 also wanted help improving the content of their presentations, such as ensuring logical flow and completeness of information. Furthermore, S2 and S4 wished for a tool that could accurately understand the nuances of both Korean and English to provide more accurate and relevant feedback.

\section{Details on Post-survey}
\label{sec:appendix_postsurvey_details}
\subsection{Post-survey Design}
In this section, we describe the design of the post-survey conducted on the 13 platform users. Students are asked to answer a 7-point Likert scale and open-ended questions related to their perceptions and experience of using the platform. The following paragraphs describe the questions used in the post-survey.

\textbf{Student’s Experience with the Platform}
\begin{enumerate}
    \item How effective was the feedback in the following areas: Vocabulary, Grammar, Content, Delivery, Organization?
    \item How effective was the feedback during partial/full rehearsal?
    \item How reliable did you find the feedback?
    \item How well did ChatGPT understand the content of your presentation?
    \item What aspects did you like or dislike about the partial/full rehearsal feedback?
    \item What was the feedback particularly helpful for?
    \item What was the feedback particularly unhelpful for?
\end{enumerate}

\textbf{Potential Learning Effects}
\begin{enumerate}
    \item To what extent did the platform improve your confidence and reduce nervousness in presenting?
    \item To what extent did the platform help enhance your vocabulary?
    \item To what extent did the platform help you identify your strengths, weaknesses, and habits in presenting?
    \item To what extent did the full rehearsal practice page resemble an online presentation environment?
    \item How useful was the platform for practicing online presentations?
\end{enumerate}

\section{Details on ChatGPT Prompts}
\label{sec:appendix_prompt_details}
This section describes the prompts used to guide ChatGPT's responses. 

Table \ref{tab:system_prompt} shows the system prompt, which includes ChatGPT's role and the context of the presentation. It also includes the summarized notes of each slide to provide ChatGPT with knowledge of the presentation content, necessary for providing accurate feedback on the content and organization components.

Table \ref{tab:feedback_prompt} shows the feedback prompt. It contains the rehearsal information, transcript, presentation rubric, and a description of the feedback style. This description guides ChatGPT in providing feedback that is specific, negative, and direct for partial rehearsals and groups common error types together for full rehearsals. The feedback prompt for the delivery component is separately designed to incorporate the speech assessment results.  

\begin{table*}[htb!]
\centering

\begin{tabularx}{\textwidth}{@{}X@{}}
\toprule
\multicolumn{1}{c}{System prompt} \\ \midrule
You are a presentation assistant for an EFL student.\\
\\
You will be given the user's presentation rehearsals, and you will provide feedback on them.\\
\\
The user is presenting at a \textless{}presentation location\textgreater{} in front of \textless{}target audience\textgreater{}.\\
\\
Here are the summarized contents of each slide of the presentation.\\
\\
\#\#\#Summarized slide contents: \textless{}Summarized slide contents\textgreater{} \\
\bottomrule
\end{tabularx}
\caption{The system prompt used to provide ChatGPT information on its role, the presentation context and slides}
\label{tab:system_prompt}
\end{table*}







\begin{table*}[htb!]
\centering
\begin{tabularx}{\textwidth}{@{}X|X@{}}
\toprule
\multicolumn{2}{c}{Feedback prompt} \\ \midrule
\multicolumn{1}{c|}{Grammar, Vocabulary, Content, Organization} &
\multicolumn{1}{c}{Delivery} \\ \midrule
\begin{tabular}[t]{@{}X@{}}The following is the transcript of a rehearsal from slide \textless{}start slide\textgreater{} to \textless{}end slide\textgreater{}:\\ \\ \#\#\#transcript: \textless{}transcript\textgreater{}\\ \\ Using the transcript, provide feedback on this rehearsal according to the following rubric:\\ \\ \#\#\#Presentation rubric: \textless{}presentation rubric\textgreater{}\\ \\ \#\#\#Feedback style: \textless{}feedback style\textgreater{}\\ \\ The final output should be in the following format:\\ \\ \#\#\#Output example: \textless{}Output example\textgreater{}\end{tabular} &
\begin{tabular}[t]{@{}X@{}}The following is the transcript of a rehearsal from slide \textless{}start slide\textgreater{} to \textless{}end slide\textgreater{}:\\ \\ \#\#\#transcript: \textless{}transcript\textgreater{}\\ \\ The following are the assessment results of the delivery of the rehearsal:\\ \\ \#\#\#Assessment results: \textless{}Delivery scores and brief explanation\textgreater{}\\ \\ The following are words that received low pronunciation scores:\\ \\ \#\#\#Poor pronunciation words: \textless{}Words with low pronunciation scores\textgreater{}\\ \\ Provide clear, concise feedback on the delivery of this rehearsal.\\ \\ The final output should be in the following format:\\ \\ \#\#\#Output example: \textless{}Output example\textgreater{}\end{tabular} \\
\bottomrule
\end{tabularx}
\caption{The feedback prompt used to guide ChatGPTs}
\label{tab:feedback_prompt}
\end{table*}

\section{Details on Presentation Rubric}
\label{sec:appendix_presentation_rubric}
Table \ref{tab:presentation_rubric} shows the presentation rubric provided to ChatGPT for feedback generation. We take the established presentation assessment criteria \cite{al2010taking, king-2002-preparing} and focus on areas where EFL students struggle, as identified through the focus group interview and existing literature \cite{hanifa2018insight, king-2002-preparing}.

\begin{table*}[ht]
\centering
\resizebox{\textwidth}{!}{%
\begin{tabularx}{\textwidth}{@{}c|X@{}}
\toprule
Criterion    & \multicolumn{1}{c}{Description}                                                                                                       \\ \midrule
Organization & \begin{tabular}[c]{@{}l@{}}- Coherent structure (Introduction, Body, Conclusion)\\ - Logical progression and transitions\end{tabular} \\ \midrule
Content &
  \begin{tabular}[c]{@{}l@{}}- Relevance to topic and objectives\\ - Appropriate depth: Comprehensive without information overload\\ - Clarity of key points and supporting arguments\end{tabular} \\ \midrule
Delivery     & \begin{tabular}[c]{@{}l@{}}- Articulation and pronunciation\\ - Pacing and timing\\ - Fluency\end{tabular}                            \\ \midrule
Grammar      & - Grammatical accuracy                                                                                                                \\ \midrule
Vocabulary   & \begin{tabular}[c]{@{}l@{}}- Contextually appropriate word choice\\ - Lexical diversity and avoidance of repetition\end{tabular}      \\ \bottomrule
\end{tabularx}
}
\caption{The presentation rubric used for feedback generation}
\label{tab:presentation_rubric}
\end{table*}


\balancecolumns
\end{document}